# On handwriting pressure normalization for interoperability of different acquisition stylus

**Marcos Faundez-Zanuy[1], Olga Brotons-Rufes[1], Carles Paul-Recarens[1], Réjean Plamondon[2]**

[1] Tecnocampus, Universitat Pompeu Fabra, Mataró, Spain

[2] Laboratoire Scribens, Département de Génie électrique, Polytechnique Montréal, Montréal, Canada

Corresponding author: Marcos Faundez-Zanuy (e-mail: faundez@tecnocampus.cat).

This work has been supported by FEDER and Ministerio de ciencia e Innovación, TEC2016-77791-C4-2-R

**ABSTRACT** In this paper, we present a pressure characterization and normalization procedure for online handwritten acquisition. Normalization process has been tested in biometric recognition experiments (identification and verification) using online signature database MCYT, which consists of the signatures from 330 users. The goal is to analyze the real mismatch scenarios where users are enrolled with one stylus and then, later on, they produce some testing samples using a different stylus model with different pressure response. Experimental results show: 1) a saturation behavior in pressure signal 2) different dynamic ranges in the different stylus studied 3) improved biometric recognition accuracy by means of pressure signal normalization as well as a performance degradation in mismatched conditions 4) interoperability between different stylus can be obtained by means of pressure normalization. Normalization produces an improvement in signature identification rates higher than 7% (absolute value) when compared with mismatched scenarios.

**INDEX TERMS** Biometrics, online handwriting, pressure sensor normalization

## I. INTRODUCTION

Function-based online signature verification relies nowadays on the acquisition of handwriting specimens using various sensors that provides position, length, velocity, acceleration, pressure, force, direction of movement, pen inclination as a function of time [1-3]. Most of these studies are run using the same set up (or a few of these) involving an instrumented stylus coupled to a passive or active writing display. The most popular systems rely on tablets, phablets, cellphones and in some cases, a finger is used instead of a stylus. For a recent overview of e-security and e-health applications based on online handwriting signals, the reader is referred to [4] as well as [5].

When the time comes to take the equipment out of the laboratory to perform large scale tests, one fundamental question arises: will the performances that have been measured in the laboratory be repeated on a large scale in an open and unconstrained environment? Two types of perturbations will come into play, behavioral and equipment perturbations. On the one hand, behavioral effects can be minimized by providing short and clear users' instructions but there are no guarantees that they will be respected. On the other hand, equipment effects can be estimated and eventually compensated with proper interoperability tests.

Among the signal functions that are used, the most critical ones are the pressure and the pen inclination because they must be directly measured with the proper sensor as compared to the other signals that can be computed from the reliable and high-resolution tablet position information. [https://www.techradar.com/reviews/pc-mac/peripherals/input-devices/graphics-tablets]

The tilt angle is not used very often because it relies on the hypothesis that the user is grabbing the stylus and signing in the same postural conditions, which is not normally true in real life applications. On the other hand, the use of pressure information in handwritten signature investigations has a long history in forensic studies where expert document analysts, working with specimens written on paper or other support, have to decide on its authenticity [6]. Using different techniques from visual inspection of the trace and its ink distribution to microscopic inspections and image processing algorithms, these experts try to recover pseudo-dynamic information regarding the way the signature has been produced. In this offline context, the pressure information stands among the most offline discriminative features.

In an online context, pressure can be measured directly from sensors embedded in the stylus, providing a continuous signal sampled at a given frequency and amplitude resolution. When the stylus is used with a tablet, x and y coordinates versus time are also available as well as tilt and azimuth angles. When the stylus is instrumented, other signals like acceleration can also be provided.

Although it has been demonstrated that there are three classes of individuals regarding pressure control, stable, unstable and piecewise time stable [7], many automatic signature verification systems use pressure information in a signal-based or feature—based approach [2]. But is the pressure signal reliable, from one device to another? To answer this question, interoperability studies must be performed and solutions to maximize interoperability must be provided if the pressure signal does not seem reliable from one device to another.

In biometrics, sensor interoperability can be defined as the capability of a recognition system to adapt to the data obtained from different sensors. Geometric interoperability refers to, for example, using writing boxes having different areas and shapes for signature acquisition and study how different geometric features are affected by these different conditions [8]. Sensor interoperability refers to the impact of using for example different tablets or stylus for data acquisition. Sensor interoperability has received limited attention in the literature and is a common challenge in biometric designs. It has been investigated for examples for iris camera detection [9], fingerprint sensors [10-11]. Regarding more specifically signature verification, the effect of using different tablets have been explored by [12] as well as by [13]. Apart from [23], to the best of our knowledge, pressure sensor interoperability has not been investigated so far, and no inter-sensor calibration model has been proposed. This is the goal of the present paper.

Biometrics systems are usually designed and trained to work with data acquired using the same sensor, which can be considered as a laboratory condition. "Real world" conditions imply signals coming from different sensors, such as different smartphone manufacturers, models, etc. As a result, changing the sensor in real operation scenarios might affect the performance of a biometric recognition system. In the case of handwriting acquisition, even using the same acquisition device, different stylus and/or stylus conditions due to usage are possible. Although x and y spatial coordinates should not be drastically affected by different stylus, this is not the case of the pressure signal. This might affect the performance of a given system.

Indeed, interoperability problems can dramatically affect the performance of biometric systems and thus, they need to be overcome. For this goal two techniques are possible:
  a)   Feature normalization
  b)   Feature selection

In this paper, we deal with pressure characterization and normalization of two different commercial stylus for the Intuos Wacom 4 tablet: the ink pen (I) able to write on paper as well as on a display for online signal acquisition and the plastic pen (P), which only acquires online information. The paper is organized as follows: section 1 is devoted to characterization of two handwriting devices. Section 2 presents the experimental results in matched and mismatched training and testing conditions (P-P, I-I, P-I, I-P). Section 3 summarizes the main conclusions of this work.

## II. CHARACTERIZATION OF HANDWRITING DEVICES

In this section, we characterize the ink pen (I) and plastic pen (P) of the Intuos Wacom 4 digitizing tablet from a physical point of view. While spatial coordinates conversion to obtain international system units is quite straightforward (conversion from pixels/tablet coordinates to millimeters), pressure requires a more laborious process, as described next.

Using a precision balance (model RADWAG WLC 10/A2) and a microscope (OPTIKA XDS-3MET), we can obtain the transfer function force applied to the stylus versus the obtained digital levels in the Intuos Wacom. In order to obtain the pressure, we need to know the surface of the pen tip and apply the formula: Pressure = Force /Surface.

It is important to take into account that the balances measure mass in grams (g) and we are interested on the weight, which is a force measured in Newtons (N). The weight of an object is a measure of the force exerted on the object by gravity, or the force needed to support it. The mass in grams is converted into weight by multiplying the mass by the earth acceleration:

Weight (N) = mass (g) * 9.807 (m/s$^2$)

Figure 1 shows the experimental setup with the precision electronic balance, whose precision is 0.1g, stylus and screw to adjust the force. Figure 2 shows the microscope used to measure the surface of the strokes produced by the tip of the pen when touching the Intuos Wacom.

The process consists of obtaining the relation between weight (in Newton) and the provided value by the tablet for each pen analyzed.

Experimental measurements with the microscope reveal the surfaces shown in table 1, where the approximate diameter of the tip of the pen has been obtained as the average of several width measurements in a straight line produced at different pressure levels. For the plastic pen, we have used a carbon copy paper in order to produce a visible line, as the plastic pen does not produce any stroke on a piece of paper. Figure 3 shows the images acquired in the microscope.

TABLE I
INK AND PLASTIC PEN NIB DIMENSIONS.

| pen | Diameter (mm) | Surface ($\pi D^2 /4$) |
|---|---|---|
| Plastic | 0.45 | 0.1590 |
| Ink | 0.319 | 0.0799 |

Pressure plots in Newton per squared millimeter (N/mm2) are shown in figure 4.

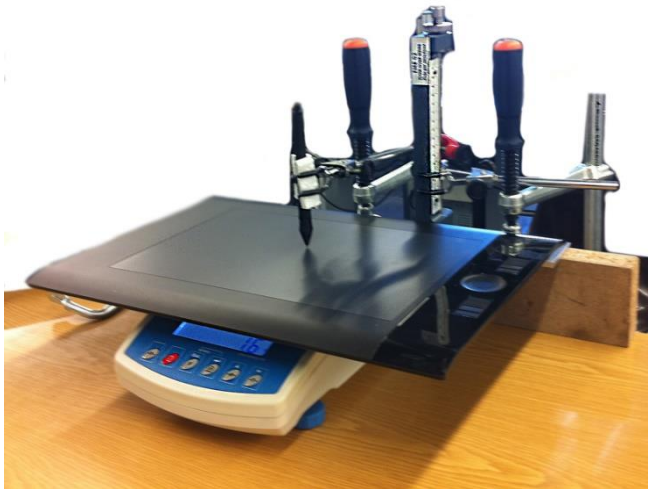

a) Frontal view

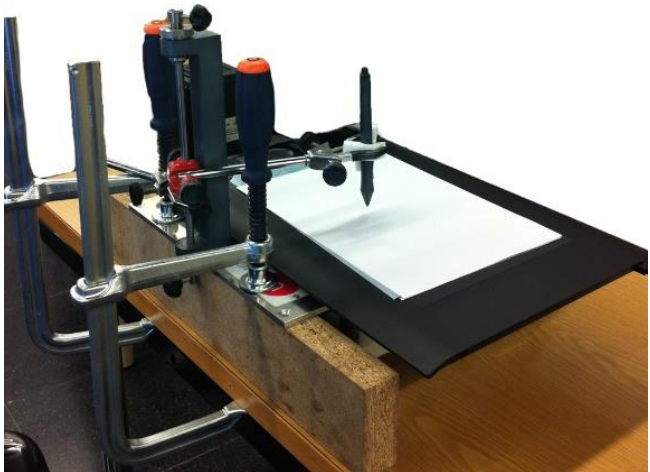

b) Rear view

**FIGURE 1. Experimental setup (precision balance RADWAG WLC 10 https://radwag.com/js/pdf_js/web/viewer.html?file=https://radwag.com/pdf2/en/WLC-Data-Sheet-EN.pdf , stylus and force actuator). a) Frontal and b) rear view**

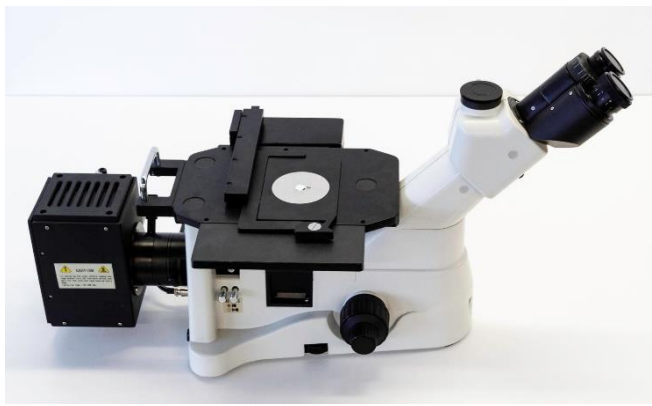

**FIGURE 2. Microscope OPTIKA XDS-3MET https://www.optikamicroscopes.com/optikamicroscopes/product/im-3-series/ to evaluate and photograph the strokes**

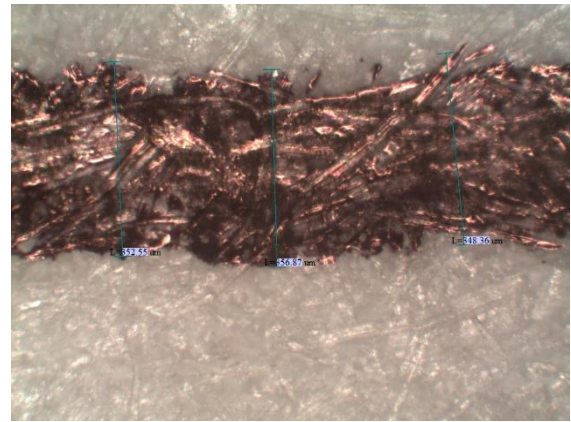

a) Ink pen at 180g weight

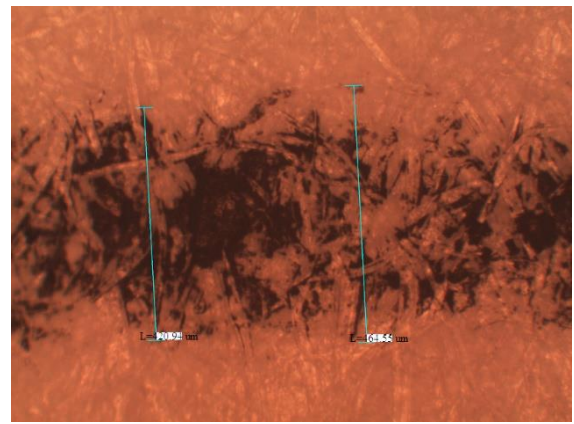

b) Plastic pen at 123g weight

FIGURE 3. Straight line and several measures from the electronic microscope for a) ink and b) plastic pen.

Figure 4 shows the setup for experimental characterization. Comparing the pressure obtained from the balance and the value obtained in the Wacom tablet, we find the experimental characteristic of the pressure sensor shown in figure 5.

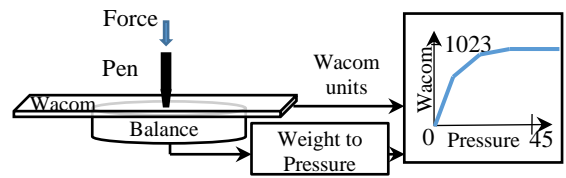

FIGURE 4. Schema for the experimental measurements to characterize the pen.

Looking at Figure 5, we observe that the plastic pen reaches saturation before the ink pen. Alternatively, the plastic pen has smaller dynamic range than the ink pen. This is not surprising since the ink pen has higher price than plastic one, it must be purchased separately, and can be considered of better quality. However, its curve also exhibits a nonlinear behavior, which affects the obtained pressure values.

Using the experimental data, we did Polynomial fitting to find the input output conversion, which is the following:

Plastic pen:
$$y_P(x) = -3,83 \cdot 10^{-5} \cdot x^6 + 0,00345 \cdot x^5 - 0,12586 \cdot x^4 + 2,4219 \cdot x^3 - 27,1975 \cdot x^2 + 199,89 \cdot x + 5,6416 \quad (1)$$
With $R^2=0.9986$, where "P" stands for plastic pen pressure

Ink pen:
$$y_I(x) = 5,48 \cdot 10^{-6} \cdot x^6 - 0,0007 \cdot x^5 + 0,03419 \cdot x^4 - 0,7571 \cdot x^3 + 6,5224 \cdot x^2 + 28,89 \cdot x - 9,90398 \quad (2)$$
With $R^2=0.9983$, where "I" stands for ink pen pressure

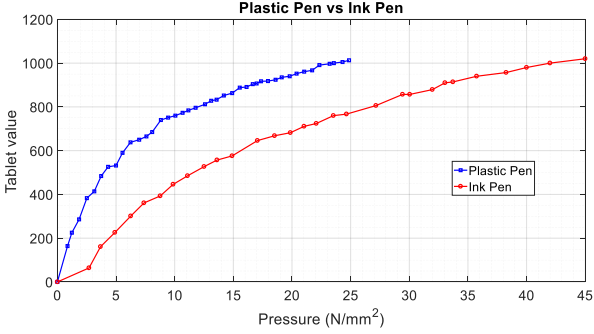

FIGURE 5. Intuos Wacom tablet value obtained for different pressure values.

The previous equations represent the coefficients for a polynomial y(x) of degree 6 that best fits (in a least-squares sense) the data in y. $R^2$ is the proportion of the variance in the dependent variable that is predictable from the independent variable. A $R^2$ of 1 indicates that the regression predictions perfectly fit the data.

## III. RELATED WORK

In the previous section, we have characterized two different stylus, a pure plastic one (P) and an ink pen (I), in order to obtain normalized pressure signals. In this section, we evaluate the relevance of pressure normalization as well as mismatch between different styluses.

### A. DESCRIPTION OF DATA BASE SETUP

One way to evaluate the interoperability of different writing devices would be to acquire data from a number of users, each one writing with different stylus. However, the intra-user variability can makes it difficult to interpret the results, due to a lack of criteria to differentiate user variability form stylus variability. Thus, since we knew the behavior of the two stylus specimens used in this study, thanks to the characterization performed in section 2, we decided to follow this evaluation procedure:
i.      Select an existing database acquired with Intuos Wacom 4 and a specific stylus. In our case the BIOSECURID, which has been obtained with an ink pen for all the acquisitions.
ii.      Normalize the pressure signal in order to obtain numerical values in an international system unit for pressure (N/mm2)
iii.      Compare experimental results with and without pressure normalization with a typical biometric signature verification algorithm.
iv.      Convert the pressure signals of the database acquired by an ink pen to those that would have been acquired by a plastic pen, according to the modules depicted in Figure 6 and evaluate the relevance of the stylus mismatch in terms of signature verification errors.

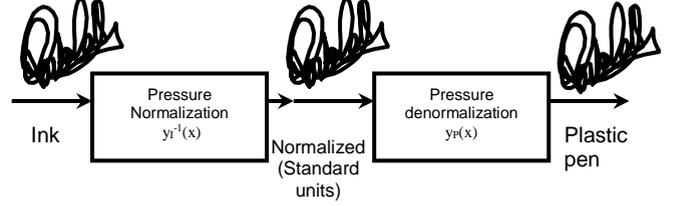

FIGURE 6. Database mapping from an ink pen pressure space to a normalized and plastic pen pressure space.

Biometric recognition is very different from most pattern recognition applications. In biometric systems a large amount of classes and a small amount of samples per class are available for training. This is the situation of biometric systems where just three to five measures per person are acquired during enrollment (for instance, user's signature, face, fingerprint, etc.). This is just the opposite situation for other pattern recognition applications where a small number of classes and a large amount of training samples are available, such as handwritten digits recognition (10 classes) for ZIP code identification. In these situations where few training samples are available the classifiers need to be necessarily simple. In fact, there is a trade-off to take into account between the model complexity and the amount of data to train the model. In addition, we focus on an application scenario were low computational burden of the recognition algorithm is desired.

Considering that the previous equations are nonlinear, an analytical inverse function $y_P^{-1}(x)$ and $y_I^{-1}(x)$ cannot be calculated. For this reason, the experimental measurements have been fitted to standard functions (see section IV.B).
Once the relevance of stylus mismatch established, two different solutions can be applied: feature selection insensitive to mismatch and feature normalization. While the first solution can be general, the second one involves the characterization of each kind of acquisition device (tablet and/or stylus), but it should be done only once for each acquisition setup (for instance, once done for a specific smartphone and stylus, it will be applicable for all the users using this same device).In a Personal Body Guard environment [14], a single user device, not doing the calibration could be a realistic scenario and even it might add a supplementary shell of protection since an impostor trying to fool the system with his own pen would be disadvantaged. However, this could be a double-edge knife strategy: if the recognition relies a lot on the stylus characteristics, then an impostor can take advantage if he stoles the stylus. On the other hand, in this scenario, protection is moving from biometrics (something you can do) to token (something you have) or a mixture of both.

Worth noting that this same interoperability problem occurs in speech/speaker recognition. In this case, training and testing signals can be acquired with different microphones. Dealing with speech signals, experimentation can be done with databases that include simultaneous speech recording using a couple of microphones, one recorded in each stereo channel [15]. This has been an active research field since several decades. See for instance [16-17]. Unfortunately simultaneous online handwriting recording cannot be done as human beings can only hold comfortably one stylus at a time. Moreover, consecutive recordings with different stylus, in addition to being time consuming, would be affected by user variability (human beings are unable to replicate the handwriting exactly in the same way).

Another option to pressure sensor empirical characterization and normalization, such as the one proposed in this paper, is blind normalization. In [18] a fully blind inversion method inspired on recent advances in source separation of nonlinear mixtures is applied to speech signals. In this case microphone saturation removal and normalization was achieved. Although blind pressure signal normalization is beyond the scope of this paper, it would be an interesting alternative approach in those scenarios where there is no access to pressure sensor empirical characterization. However stylus characterization and normalization in online handwritten analysis is more straightforward than microphone characterization for speech signal analysis: speech channel is noisy due to ambient noise, distance from head to microphone, etc. and handwriting is "cleaner" as there is no ambient noise effect and signal is always acquired "on surface", and thus, always at the same distance (height).

## IV. EXPERIMENTAL RESULTS

In this section, we present the experimental results using the data base setup described in section III.

### A. EXPERIMENTAL CONDITIONS

In this paper, we have used the same experimental conditions and algorithms than in our previous paper [19], which is an improvement of Vector Quantization algorithm applied to online signature recognition [20]. We selected this algorithm because it is very fast and thus suitable for intensive experimentation. Worth to mention that we performed relative comparisons in mismatch conditions, always with the same algorithm, and the use of a most powerful recognition algorithm should not significantly have affected the conclusions, unless some indirect robust feature selection is incorporated in the training of the classifier.

The multi-section codebook model approach consists of splitting the training samples into several sections. For example, when using three sections, each signature is split into three equal length parts (initial, middle, and final sections). In this case, three codebooks are generated for each user, each codebook being adapted to one portion of the signature. We will use a codebook size ranging from 1 to 8 bits per section (thus, each codebook section consists of $2^1=2$ to $2^8=256$ vectors). Final decision is taken by combining individual contributions of each section by simple averaging.

If the database contains P users and the model consists of S sections, we will have S codebooks for each person. In this paper we have considered S=2 sections. Thus, we had two codebooks per person. Experimental results were obtained with the MCYT database [21], acquired with a WACOM graphics tablet. The sampling frequency for signal acquisition is set to 100 Hz, yielding the following set of information for each sampling instant:

1. Position along the x-axis, $x_t$ : [0–12 700], corresponding to 0–127 mm;
2. Position along the y-axis, $y_t$ : [0–9700], corresponding to 0–97 mm;
3. Pressure $p_t$ applied by the pen: [0–1024];
4. Azimuth angle $\gamma_t$ of the pen with respect to the tablet: [0–3600], corresponding to 0–360º;
5. Altitude angle $\varphi_t$ of the pen with respect to the tablet: [300–900], corresponding to 30–90º;

We have used feature vectors composed of these five measurements. We analyzed the signatures of 330 different users. Each target user produces 25 genuine signatures in different acquisition sessions.

Training and testing signatures have been chosen in the following way:

- We performed identification experiments, using the first 5 signatures per person for training and 5 different signatures per person for testing (signatures 6–10, acquired in a different acquisition session). This implies a total number of $330 \times 330 \times 5$ tests.
- We performed verification experiments, using the first 5 signatures per person for training and 5 different genuine signatures per person for testing (signatures 6–10). In addition, we performed $330 \times 329 \times 5$ impostor tests (random forgeries).
- We performed verification experiments, using the 25 skilled forgeries available in MCYT, produced by five other users trying to imitate the attacked signature. This implies a total number of $330 \times 5$ genuine tests plus $330 \times 25$ impostor tests (skilled forgeries).

These experimental conditions provide with 95% confidence the statistical significance in experiments with an empirical error rate, down to [20] $\hat{P} = 0.02\%$ for random forgeries and $\hat{P} = 1.01\%$ for skilled forgeries.

Verification systems can be evaluated using the False Acceptance Rate (FAR, those situations where an impostor is accepted) and the False Rejection Rate (FRR, those situations where a genuine user is incorrectly rejected). A trade-off between both errors (FA = False Acceptance and FR = False Rejection) usually has to be established by adjusting a decision threshold. The performance can be plotted in a Receiver

Operator Characteristic (ROC) or in a Detection Error Trade-off (DET) plot [22].

We have used a single point of the DET plot for comparison purposes: the minimum value of the Detection Cost Function (DCF). This parameter is defined as [21]:

$DCF = C_{FR} \times P_{FR} \times P_{true} + C_{FA} \times P_{FA} \times P_{false}$

where $C_{FR}$ is the cost of a rejection of a true specimen, $C_{FA}$ is the cost of a false acceptance, $P_{FR}$ is the probability of a false rejection, $P_{FA}$ is the probability of a false acceptance, $P_{true}$ is the a priori probability of the target, and $P_{false} = 1 - P_{true}$. Here we used equal costs: $C_{FR} = C_{FA} = 1$. In addition, we also provided the complete DET plots and EER line.

### B. FITTING CURVES

The standard curves have been adjusted to the experimental points by least squares fitting, using Matlab®'s optimization toolbox function "lsqcurvefit". Despite polynomials from expressions (1) and (2) were already valid fitting curves for the available experimental measurements, they have six or seven parameters and they do not cross over the origin, so they had to be modified and it was ponderous to work with them. Eventually, different curves with different properties were tested: they had to be monotonically increasing and concave (for the de-normalization curve), they had to have a small number of parameters and had to include the origin in their domain. After several tries, the best fits have been obtained with a logarithmic and an elliptic curve for the ink and the plastic pen, respectively.

Besides, the curves have been rescaled such that their output ranges from 0 to 1024, the same range of the digitizing tablet. This rescaling is important, otherwise the weight of the pressure signal is reduced and recognition rates are lower. This process can be seen as weighted distance computation where pressure signal is emphasized (3):

$$d(\vec{a},\vec{b}) = \left(\sum_{i=\{x,y,p,\gamma,\varphi\}}[\omega_i(x_i - y_i)]^2\right)^{\frac{1}{2}} \quad (3)$$

where the weighting for pressure component is $\omega_p = \frac{1024}{45}$ for the ink pen and $\omega_p = \frac{1024}{25}$ for the plastic pen (according to Figure 5, 45 and 25 are the maximum pressure values provided by the stylus), and the other weighting factors are: $\omega_x = \omega_y = \omega_\gamma = \omega_\varphi = 1$.

In Figure 5, we observed that plastic pen reaches saturation earlier than ink pen. Thus, we can consider that ink pen has a better pressure sensor than the plastic one.

The ink pen measurements have been fitted with:

$$p(w) = F_i \cdot \frac{-1}{A_2} \cdot \ln\left(1 - w/A_1\right) \quad (4)$$

Where:
$A_1 = 1148.6344 \; ; \; A_2 = 0.0468 \; ; \; F_i = 21.5761$

We chose this function since it can be analytically inverted and it fits quite well the experimental curve obtained by direct measurements.

While equation (4) is the function that normalizes the pressure data (from the digitizing tablet pressure level ("w") to standard pressure units ("p", in MPa), the inverse function, here after called the denormalization function, is:

$$w(p) = A_1 \cdot \left(1 - e^{-A_2 \cdot p/F_i}\right) \quad (5)$$

Regarding the plastic pen, the normalization (6) and denormalization (7) functions are:

$$p(w) = F \cdot \left[R_1 - \sqrt{R_1^2 - (w/R_2)^2}\right] \quad (6)$$

$$w(p) = R_2 \cdot \sqrt{2R_1 \cdot p/F - (p/F)^2} \quad (7)$$

Respectively, where:
$R_1 = 33{,}5234 \; ; \; R_2 = 31{,}1303 \; ; \; F = 37.8450$

With these functions to obtain $R^2=0.991$ for plastic pen and $R^2=0.995$ for the ink pen

### C. TEST SCENARIOS

Table 2 describes the tested situations. The original test results (scenario 1) are included in order to compare to the new ones. Besides, there are two sets of tests: the match (scenarios 2 and 3), and the mismatch situation (scenarios 4 to 7). The last one has been realized both using raw data (scenarios 4 and 5) and normalizing these, in order to correct the mismatch effect (scenarios 6 and 7).

TABLE II
TEST SCENARIOS DESCRIPTION

|  | TRAIN | TEST | SCENARIO |
|---|---|---|---|
| ORIGINAL TEST (RAW DATA) | I (INK) | I (INK) | 1 |
| MATCH (NORMALIZED DATA) | I (INK) | I (INK) | 2 |
|  | P (PLASTIC) | P (PLASTIC) | 3 |
| MISMATCH (RAW DATA) | P (PLASTIC) | I (INK) | 4 |
|  | I (INK) | P (PLASTIC) | 5 |
| MISMATCH (NORMALIZED DATA) | P (PLASTIC) | I (INK) | 6 |
|  | I (INK) | P (PLASTIC) | 7 |

### D. EXPERIMENTAL RESULTS

Match Test – random forgeries

Figure 7 shows the match test (scenarios 1 to 3) in an identification experiment. We observe that normalized ink data (scenario 2) and synthetically generated and normalized plastic pen data (scenario 3) yield almost the same results. This shows that normalization process is not degrading identification accuracies. In some conditions, it is even slightly better.

Figure 8 shows the match test (scenarios 1 to 3) in a verification experiment. We observe a slight degradation in Detection Cost Function, around a 10%, especially when the model size is small (≤ 5 bits per section). It is worth mentioning that in order to simplify the analysis and restrict it to pressure normalization, we performed direct verification without any normalization such as cohort or universal background model, which indeed would improve the results in all the scenarios.

Figure 9 shows the DET plot for the same configurations of figure 8. In this case the model size is 7 bits per section. The intersection between EER line and DET plot indicates the EER while the circle on the DET plot indicates the minDCF value for the specific setup.

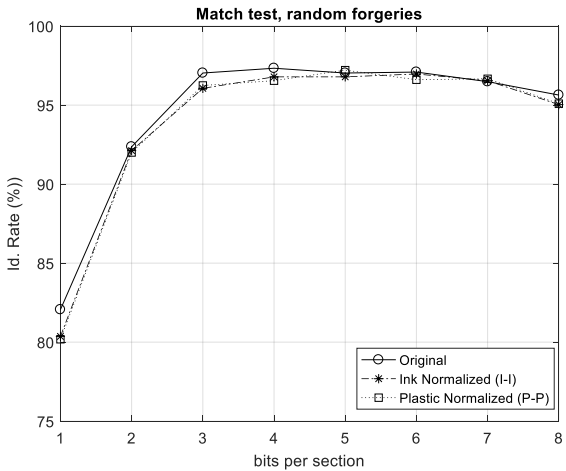

FIGURE 7. Identification rates vs bits per section in matched conditions.

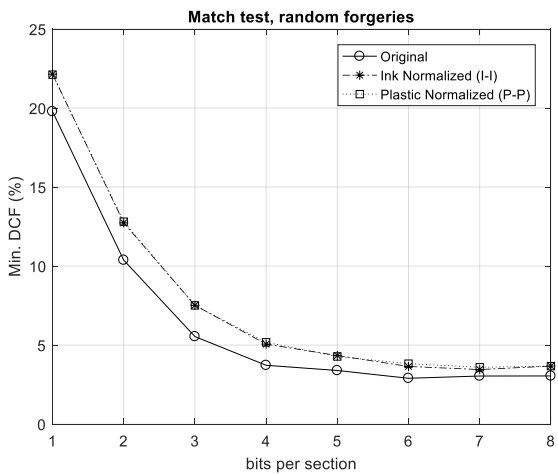

FIGURE 8. Verification error (min DCF) vs bits per section in matched conditions and random forgeries.

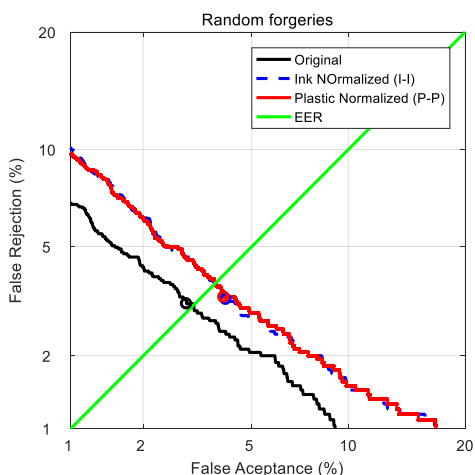

FIGURE 9. Detection Error Tradeoff (DET) plots for matched experiments, random forgeries, with a model size of 7 bits. The circles indicate the minimum Detection Cost function (min DCF)

Mismatch tests – random forgeries

Figure 10 shows the mismatch test (scenarios 4 to 7) in an identification experiment. We observe that non-normalized experiments did not reach satisfactory results. This situation corresponds to the real scenario where users are enrolled in one digitizing device and testing sample comes from a different device. We observe that the mismatch in the pressure signal handicaps the accuracies and that pressure normalization is mandatory for good accuracies. In fact, it is better to discard pressure signal rather than using non-matched pressure signal. Figure 10 includes the "no pressure data" experimental results, which is below normalization strategy but, sometimes, over mismatched results (P-I, I-P, where first letter corresponds to train condition and second one to testing condition).

Figure 11 shows the mismatch test (scenarios 4 to 7) in a verification experiment. We observe a degradation in Detection Cost Function for codebook sizes ≥ 6. Figure 11 includes the "no pressure data" experimental results, which are always worse than mismatched results (P-I, I-P, where first letter corresponds to train condition and second one to testing condition), and normalized results.

A more realistic mismatch situation is where at the enrolment step all the users use the same tablet and pen but as time goes on, some of them keep the same enrolment acquisition system and some of them update to a new one with different characteristics. In figure 12 we represent the signature identification rates as function of the percentage of users in mismatch conditions. 0% means that all the users perform the test with the same testing pen that was used for training. Figure 13 represents the verification errors in this same conditions.

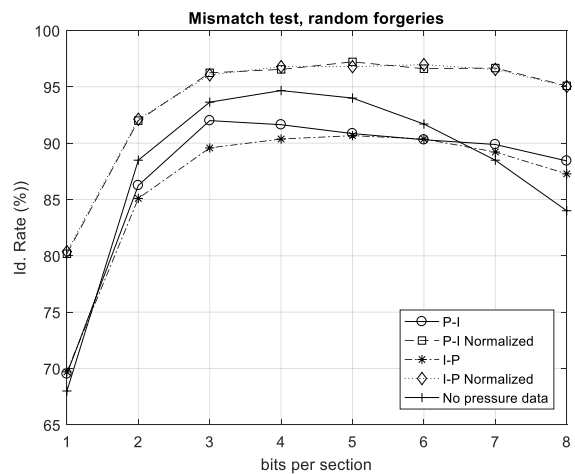

FIGURE 10. Identification rates vs bits per section in mismatched conditions.

We can see that global signature identification rates and verification errors are worse as the number of users in mismatch conditions grows up. The worse case situation corresponds to 100% users performing the test with a different pen.

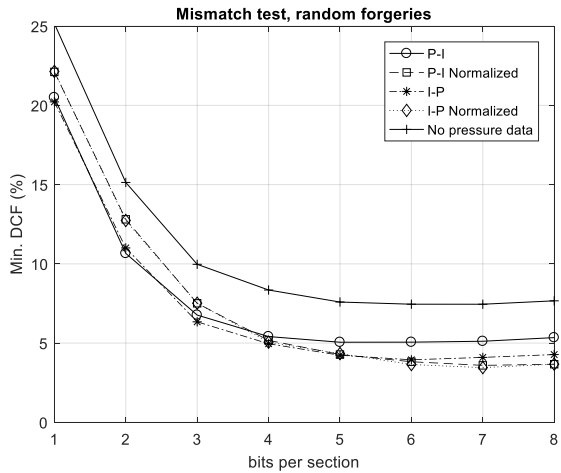

FIGURE 11. Verification error (min DCF) vs bits per section in mismatched conditions and random forgeries.

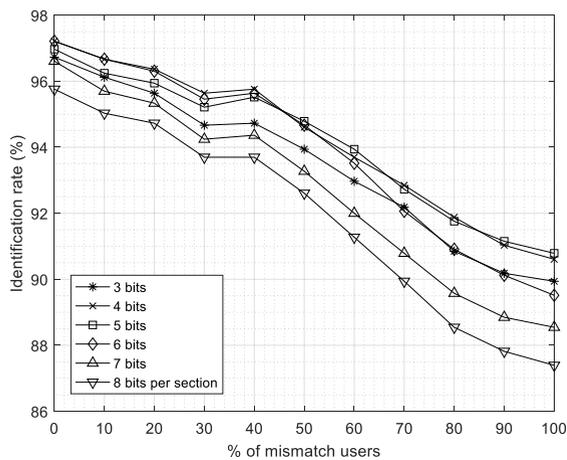

FIGURE 12. Identification rates vs % of users in mismatch between training and testing pen

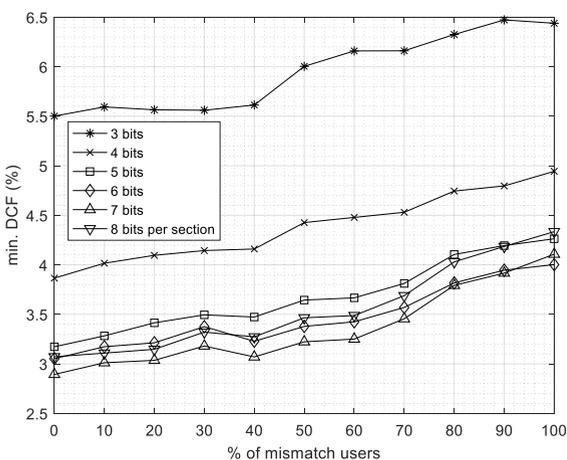

FIGURE 13. Verification error vs % of users in mismatch between training and testing pen, using random forgeries.

Figure 14 shows the DET plot for the same configurations of figure 11. In this case the model size is 7 bits per section. The intersection between EER line and DET plot indicates the EER while the circle on the DET plot indicates the min DCF value for the specific setup.

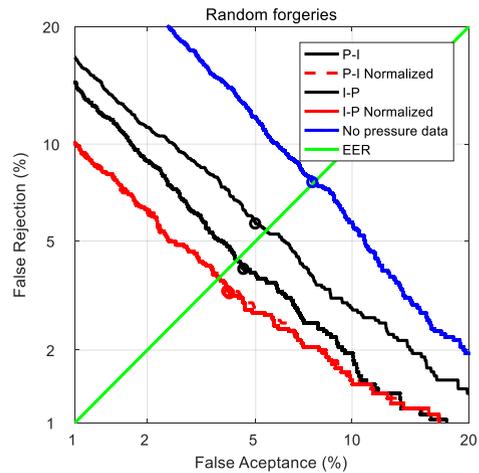

FIGURE 14. DET plots for matched experiments, with a model size of 7 bits. The circles indicate the min DCF

In figure 14 we can observe, on the one hand, almost identical performance for normalized pressure in mismatch conditions. On the other hand un-normalized tests in mismatch condition show a significant degradation, although accuracies are better than removing the pressure signal at all.

Match Test – skilled forgeries

Figure 15 is analogous to figure 8 but using skilled forgeries attacks. As expected, the errors rise when compared to random forgeries by a factor of 2. We observe too that pressure normalization does slightly degrade the accuracy (<1% for 7 bit model).

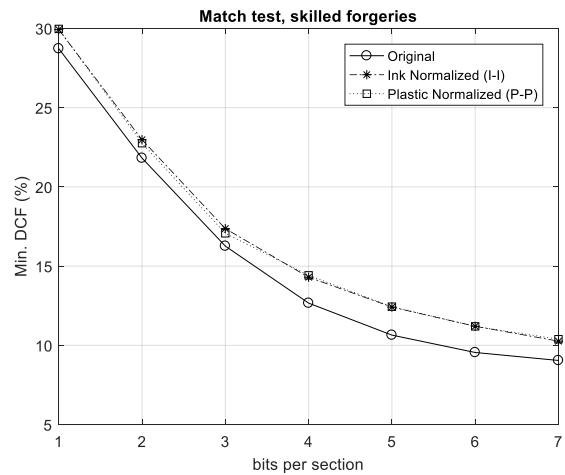

FIGURE 15. Verification error (min DCF) vs bits per section in matched conditions and skilled forgeries

Figure 16 is analogous to figure 9 but using skilled forgeries attacks. As expected, the errors rise when compared to random forgeries. We observe too that pressure normalization does not degrade the accuracy.

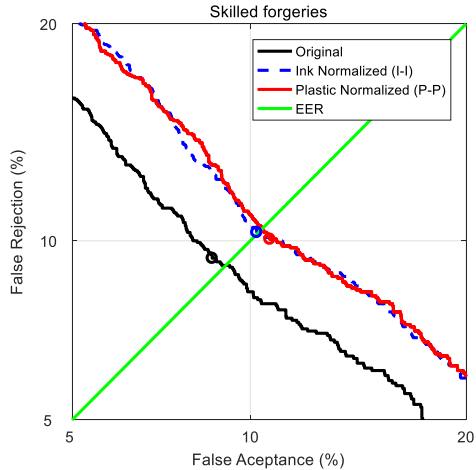

FIGURE 16. DET plots for matched experiments, skilled forgeries, with a model size of 7 bits. The circles indicate the min DCF

Mismatch tests – skilled forgeries

Figure 17 is analogous to figure 11 but using skilled forgeries attacks. As expected, the errors rise as compared to random forgeries. We observe better performance when using pressure normalization and the worse performance if pressure data is discarded.

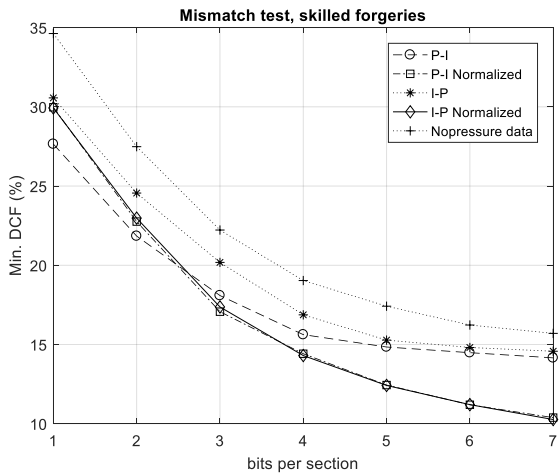

FIGURE 17. Verification error (min DCF) vs bits per section in mismatched conditions and skilled forgeries

Figure 18 shows the DET plot for mismatch scenarios and skilled forgeries. In this case the model size is 7 bits per section. The intersection between EER line and DET plot indicates the EER while the circle on the DET plot indicates the min DCF value for the specific setup.

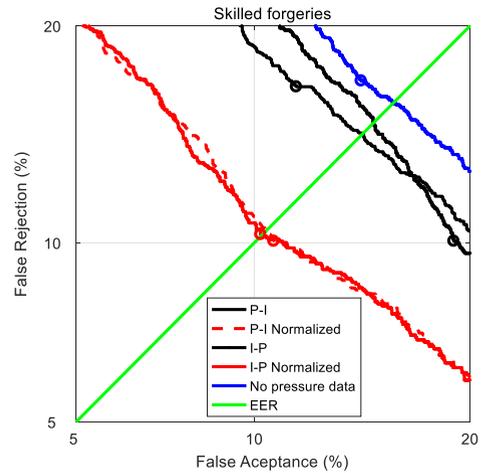

FIGURE 18. DET plots for mismatched experiments, skilled forgeries, with a model size of 7 bits. The circles indicate the min DCF

Table 2 summarizes the minDCF and EER of all the experiments, with a model size of 7 bits.

TABLE III
SUMMARY OF VERIFICATION EXPERIMENTS

| SCENARIO | FORGERY | MINDCF (%) | EER (%) |
|---|---|---|---|
| 1 (ORIGINAL) | RANDOM | 3.04 | 3.08 |
| 2 (I-I NORMALIZED) | RANDOM | 3.67 | 3.72 |
| 3 (P-P NORMALIZED) | RANDOM | 3.67 | 3.75 |
| 4 (P-I) | RANDOM | 5.12 | 5.52 |
| 5 (I-P) | RANDOM | 4.28 | 4.3 |
| 6 (P-I NORMALIZED) | RANDOM | 3.67 | 3.75 |
| 7 (I-P NORMALIZED) | RANDOM | 3.67 | 3.72 |
| NO PRESSURE | RANDOM | 7.67 | 7.85 |
| 1 (ORIGINAL) | SKILLED | 9.05 | 9.13 |
| 2 (I-I NORMALIZED) | SKILLED | 10.27 | 10.43 |
| 3 (P-P NORMALIZED) | SKILLED | 10.39 | 10.39 |
| 4 (P-I) | SKILLED | 14.16 | 14.36 |
| 5 (I-P) | SKILLED | 14.58 | 14.81 |
| 6 (P-I NORMALIZED) | SKILLED | 10.39 | 10.39 |
| 7 (I-P NORMALIZED) | SKILLED | 10.27 | 10.43 |
| NO PRESSURE | SKILLED | 15.7 | 16.1 |

## V. CONCLUSIONS

In this paper, we have presented a set of experimentations for pressure normalization as well as experiments comparing the match and mismatch situation of the stylus, which provides a different pressure response. The evaluation has been done with a multi-session database which consists of 330 users. The main conclusions are:

- The stylus pressure response is far from linear, with an important saturation effect for high pressure values. This shape can be properly fitted with an exponential curve, which is easier to invert than a polynomial one.
- The Intuos Wacom plastic pen reaches saturation for lower pressure than the Wacom ink pen, which presents a higher dynamic range (almost two times higher) and thus, it is has a better pressure sensor.

- Although simultaneous manual acquisition with different stylus is not possible due to intra-user variability, we proposed a procedure to obtain one stylus signal from another one thanks to a normalization and de-normalization procedure.
- We observed a reduction in identification rates in mismatch condition, which is higher than 7% absolute identification rate (96.85% identification rate with proposed normalized pressure versus 89.52% identification rate in mismatch scenario, where these numbers have been obtained from figure 10 for 6 bits, which corresponds to the highest identification rate). We observe a reduction in verification errors in mismatch condition and random forgeries, which is 1.8% (3.4% DCF with normalized pressure versus 5.2% DCF in mismatch scenario, where these numbers have been obtained from figure 11 with 7 bits, which provides the minimum DCF). We observe a reduction in verification errors in mismatch condition and skilled forgeries, which is around 4.5% in most of the experiments. Worth to mention that in order to simplify the analysis and restrict it to pressure normalization we performed direct verification without any normalization such as cohort or universal background model, which indeed would improve the results in all the scenarios.
- Pressure normalization can manage the real scenario where some users write using the same stylus used for training and some others have a new one with different characteristic. The unique required process is a characterization and normalization of the stylus, which has been described in this paper. This is a fast process as only one additional equation must be applied to pressure signal.
- In pressure sensor mismatch scenarios better biometric identification rates are obtained discarding the pressure signal, unless normalization strategy is applied. However, in signature verification application, discarding pressure signal provides worse results than matched/mismatched conditions.
- Pressure normalization has not been applied so far in existing scientific literature, and it is a key point for obtaining good recognition accuracies in interoperability scenarios.

As the technological evolution produces new devices, which exhibit new technical characteristics, it is worth to facilitate the interoperability between these devices. Notice that some of the largest existing databases were acquired in the past with discontinued products (see for instance the list of discontinued Wacom tablets at https://www.wacom.com/en-us/for-business/products/discontinued-products ). To the best of our knowledge, this is the first paper devoted to facilitate the interoperability of pressure values obtained from different digitizing tablets.

The key point is that some commercial devices for online handwritten signal acquisition are already in the market. They can be used for signature, text and drawings (e-security and e-health applications). The pressure signal provided by these devices is the analog to digital value directly provided by the Analog to Digital Converter. Thus, direct comparison between signals acquired by different devices often results in low recognition rates, as demonstrated in this paper, and some normalization process is mandatory to rise these recognition performances. For instance, this can be done using the technical methodology presented in this paper, which has been satisfactorily tested with experimental data.

**REFERENCES**

[1] Plamondon R, Pirlo G, Impedovo D. (2014). On-line Signature Verification. In Doermann D, Tombre K. (EDS) *Handbook of Document Image Processing and Recognition*. Springer, Germany, pp 917-947.

[2] Diaz, M., Ferrer, M. A., Impedovo, D., Malik, M.I., Pirlo, G., Plamondon, R. (2019). A perspective analysis of handwritten signature technology. . ACM Computing Surveys. 51(6): 39 pages.D. Impedovo, G. Pirlo (2018) Automatic signature verification in the mobile cloud scenario: survey and way ahead, IEEE Transactions on Emerging Topics in Computing. doi: 10.1109/TETC.2018.2865345

[3] D. Impedovo, G. Pirlo (2018) Automatic signature verification in the mobile cloud scenario: survey and way ahead, IEEE Transactions on Emerging Topics in Computing. doi: 10.1109/TETC.2018.2865345

[4] Faundez-Zanuy, M., Fierrez, J., Ferrer, M.A. and Plamondon, R. Handwriting Biometrics: Applications and Future Trends in e-Security and e-Health. Cogn Comput 12, 940–953 (2020). https://doi.org/10.1007/s12559-020-09755-z

[5] Plamondon, R., Marcelli, A., Ferrer, M.A., (Eds) (2020). The Lognormality Principle and its Applications for e-Security, e-Learning and e-Health, World Scientific Publishing, New Jersey, London, Singapore, 415 pages.

[6] ASTM Standard E444, (2009) Standard Guide for Scope of Work of Forensic Document Examiners, ASTM International, West Conshohocken, PA, 2009, DOI: 10.1520/Eo444-09

[7] Schomaker, L.R.B., Plamondon, R. The relation between pen force and pen-point kinematics in handwriting. Biol. Cybern. 63, 277–289 (1990). https://doi.org/10.1007/BF00203451

[8] Giuseppe Pirlo, Fabrizio Rizzi, Annalisa Vacca, and Donato Impedovo (2015) Interoperability of Biometric Systems: Analysis of Geometric Characteristics of Handwritten Signatures. Proc. International Conference on Image Analysis and ProcessingICIAP 2015: New Trends in Image Analysis and Processing -- ICIAP 2015 Workshops pp 242-249

[9] Sunpreet S. Arora, Mayank Vatsa, Richa Singh and Anil Jain "Iris recognition under alcohol influence: A preliminary study," 2012 5th IAPR International Conference on Biometrics (ICB), New Delhi, 2012, pp. 336-341, doi: 10.1109/ICB.2012.6199829

[10] ] Ross, A., Jain, A. "Biometric sensor interoperability: A case study in fingerprints". In: Maltoni, D., Jain, A.K. (eds.) BioAW 2004. LNCS, vol. 3087, pp. 134–145. Springer, Heidelberg (2004)

[11] Luca Lugini, "Interoperability of fingerprint sensors and matching algorithms" (2014). Graduate Theses, Dissertations, and Problem Reports. 7328. https://researchrepository.wvu.edu/etd/7328

[12] Alonso-Fernandez F., Fierrez-Aguilar J., Ortega-Garcia J. (2005) Sensor Interoperability and Fusion in Signature Verification: A Case Study Using Tablet PC. In: Li S.Z., Sun Z., Tan T., Pankanti S., Chollet G., Zhang D. (eds) Advances in Biometric Person Authentication. IWBRS 2005. Lecture Notes in Computer Science, vol 3781. Springer, Berlin, Heidelberg. https://doi.org/10.1007/11569947_23

[13] Impedovo, D., Pirlo, G., Sarcinella, L., Vessio, G. "An Evolutionary Approach to address Interoperability Issues in Multi-Device Signature Verification", 2019 IEEE International Conference on Systems, Man and Cybernetics (SMC) October 2019 Pages 3048–3053 https://doi.org/10.1109/SMC.2019.8914523

[14] Plamondon R, Pirlo G, Anquetil E, Rémi C, Teuling H-L, Nakagawa M. (2018). Personal digital bodyguards for e-security, e-learning and e-health: A prospective survey. Pattern Recognition. 81: 633-659.

[15] C. Alonso-Martinez and M. Faundez-Zanuy, "Speaker identification in mismatch training and testing conditions," 2000 IEEE International Conference on Acoustics, Speech, and Signal Processing. Proceedings (Cat. No.00CH37100), Istanbul, Turkey, 2000, pp. II1181-II1184 vol.2, doi: 10.1109/ICASSP.2000.859176.

[16] Richard J. Mammone, Xiaoyu Zhang, Ravi P. Ramachandran. "Robust speaker recognition: A feature-based approach". October 1996, IEEE Signal Processing Magazine 13(5):58, DOI: 10.1109/79.536825

[17] Sharada V. Chougule. Mahesh S. Chavan "Robust Spectral Features for Automatic Speaker Recognition in Mismatch Condition" Procedia Computer Science, Volume 58, 2015, Pages 272-279. https://doi.org/10.1016/j.procs.2015.08.021

[18] Solé-Casals, J., Faúndez M. (2006). Application of the mutual information minimization to speaker recognition/identification improvement. Neurocomputing, 69, Issues 13-15, pp. 1467-1474. https://doi.org/10.1016/j.neucom.2005.12.023

[19] Marcos Faundez-Zanuy and Juan Manuel Pascual-Gaspar "Efficient On-line signature recognition based on Multi-section VQ" Pattern Analysis and Applications. Volume 14, Number 1 pp. 37-45, February 2011

[20] Marcos Faundez-Zanuy "On-line signature recognition based on VQ-DTW". Pattern Recognition Vol. 40 (2007) pp.981-992 https://doi.org/10.1016/j.patcog.2006.06.007

[21] J. Ortega-Garcia, J. Fierrez-Aguilar, D. Simon, M. Faundez, J. Gonzalez, V. Espinosa, A. Satue, I. Hernaez, J. J. Igarza, C. Vivaracho, D. Escudero and Q. I. Moro, "MCYT baseline corpus: A bimodal biometric database", IEE Proceedings Vision, Image and Signal Processing, Vol. 150, n. 6, pp. 395-401, December 2003

[22] A. Martin, G. Doddington, T. Kamm, M. Ordowski, Mark A. Przybocki (1997) "The DET curve in assessment of detection performance". In: European speech processing conference Eurospeech, vol 4, pp 1895–1898

[23] Rubén Tolosana, Rubén Vera-Rodríguez, Javier Ortega-Garcia, Julian Fiérrez "Preprocessing and Feature Selection for Improved Sensor Interoperability in Online Biometric Signature Verification". Published in IEEE Access 2015. DOI:10.1109/ACCESS.2015.2431493

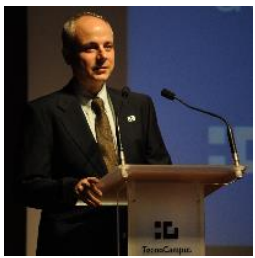

**MARCOS FAUNDEZ-ZANUY** was born in Barcelona, Spain. He received the PhD degree in 1998 at the Polytechnic University of Catalunya. He is now Full Professor at ESUP Tecnocampus Mataro and heads the Signal Processing Group there. From 2009 till 2018 he was the dean of Escola Superior Politecnica Tecnocampus (Polytechnic University of Catalunya till 2015, Pompeu Fabra University after 2015). Since 2010 till 2019 he was the head of research at Tecnocampus. His research interests lie in the fields of biometrics applied to security and health. He was the initiator and Chairman of the European COST action 277 Nonlinear speech processing, and the secretary of COST action 2102 Cross-Modal Analysis of Verbal and Non-Verbal Communication. He is an author of more than 100 papers indexed in WoS, more than 100 conference papers, around 10 books, and responsible of 10 national and European research projects. Prof. Faundez-Zanuy is a Spanish liaison of EURASIP.

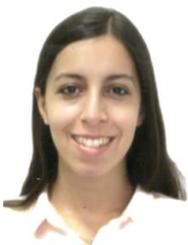

**OLGA BROTONS-RUFES** was born in Barcelona, Spain. She got the BEng Degree in Industrial and Automatization Electronic Engineering in 2020, as well as the BEng Degree in Mechanical Engineering, at ESUPT Tecnocampus, Pompeu Fabra University (Mataró, Spain). Her first research project was her final work for her Degree in Electronics, which was developed under Prof. Marcos Faundez-Zanuy direction. Currently, she is working in a R&D project related to water infrastructures at Aigües de Barcelona, which is a water company that was founded in 1867. The company collects, purifies, transports and distributes water to other companies and to residents.

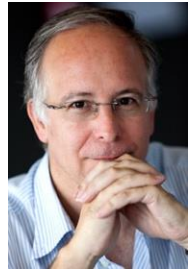

**CARLES PAUL-RECARENS** was born in Arenys de Munt, Spain. He received a B.Sc degree in Physics from the University of Barcelona. Then he got a Master's Degree in Physics and Mathematics from the Polytechnic University of Catalunya and a Master's Degree in History of Science: Science, History and Society from the Autonomous University of Barcelona. He is now a Lecturer at ESUP Tecnocampus Mataró. Since 2011 he has been an Eurostars Technical Expert with identification number 4507. He is currently the Scientific Director of INNOVEM and SIGMA where he is also the founding partner of both companies. At Innovem he created the DeltaQon biomechanical assessment system, composed of inertial sensors to evaluate the pathologies that affect the functionality of the musculoskeletal system. At Sigma he is working with predictive maintenance for electrical machines. He is also pursuing a Ph.D degree in Electrical Engineering with a new type of motors on the basis of high currents without external magnetic fields.

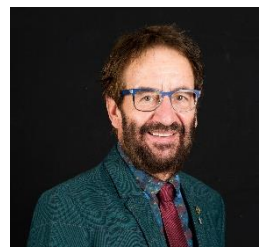

**RÉJEAN PLAMONDON** received a B.Sc. degree in Physics, and M.Sc.A. and Ph.D. degrees in Electrical Engineering from Université Laval, Québec, Canada. Then he joined the faculty of the École Polytechnique, Montréal, Canada, where he is currently a Full Professor. He has been Head of the Department of Electrical and Computer Engineering from 1996 to 1998 and President of École Polytechnique from 1998 to 2002. During all these years at Polytechnique, he has been the Head of Laboratoire Scribens.

Throughout his career, Professor Plamondon has been working in Pattern Recognition and Artificial Intelligence, particularly on the study of emerging phenomena and behavior in biological and physical systems exploiting various convergence theorems. He is a leader in the fields of analysis and processing of on-line and off-line handwriting. Studying and modeling the emergent properties of neuromotor systems involved in the generation of human motions, he proposed innovative solutions to technical problems for the design of automated systems dedicated to signature verification, gesture and handwriting processing, interactive tools to help children learn to write. He also developed robust methods for the analysis and interpretation of neuromuscular information from movement kinematic signals to characterize the fine motor skills in healthy persons or patients suffering from various diseases. His main contribution is the development of the Kinematic Theory of human movements and its Lognormality Principle, which have been successfully used to describe the central and peripheral features of fingers, wrists, arms, head, trunk, eye movements as well as of speech. He has investigated these bio signals extensively to develop creative and powerful methods and applications in various domains of engineering and social sciences.

Full member of the Canadian Association of Physicists, the Ordre des Ingénieurs du Québec, the Union Nationale des Écrivains du Québec, Dr. Plamondon is also an active member of several international societies. He is a lifelong Fellow of the Institute of Electrical and Electronics Engineers (IEEE; 2015); Fellow of the Netherlands Institute for Advanced Study in the Humanities and Social Sciences (NIAS; 1989), of the International Association for Pattern Recognition (IAPR FELLOW; 1994) and of the Institute of Electrical and Electronics Engineers (IEEE FELLOW; 2000). He has been involved in the planning and organization of numerous international conferences and workshops and has worked with scientists from many countries all over the world. He is the author or co-author of more than 400 publications and owner of four patents. He has edited or co-edited seven books and several Special Issues of scientific journals. He has also published a children's book, a short story and four collections of poems. He recently received the IAPR/ICDAR 2013 outstanding achievement award "for theoretical contributions to the understanding of human movement and its applications to signature verification, handwriting recognition, instruction, and health assessment, and for promoting on-line document processing in numerous multidisciplinary fields."